\documentclass[10pt,twocolumn,aps,prl,notitlepage,showpacs,floatfix,amsmath,amssymb;twocolumn]{revtex4-1}
\usepackage[latin9]{inputenc}
\usepackage{amsmath}
\usepackage{graphicx}
\usepackage{esint}

\makeatletter

\usepackage{graphicx}
\usepackage{float}

\usepackage{epstopdf}
 \usepackage{setspace}

\newcommand{\scS}{\mathcal{S}}

\newcommand{\twovec}[3][2]{\left(\begin{array}{c} #2 \\ #3 \end{array} \right)}

\makeatother

\begin{document}

\title{Beyond Gaussian pair fluctuation theory for strongly interacting
Fermi gases}

\author{Brendan C. Mulkerin, Xia-Ji Liu, and Hui Hu}

\affiliation{Centre for Quantum and Optical Science, Swinburne University of Technology,
Melbourne 3122, Australia.}

\date{\today}
\begin{abstract}
Interacting Fermi systems in the strongly correlated regime play a
fundamental role in many areas of physics and are of particular interest
to the condensed matter community. Though weakly interacting fermions
are understood, strongly correlated fermions are difficult to describe
theoretically as there is no small interaction parameter to expand
about. Existing strong coupling theories rely heavily on the so-called
many-body $T$-matrix approximation that sums ladder-type Feynman
diagrams. Here, by acknowledging the fact that the effective interparticle
interaction (i.e., the vertex function) becomes smaller above three
dimensions, we propose an alternative way to reorganize Feynman diagrams
and develop a theoretical framework for interacting Fermi gases beyond the
ladder approximation. As an application, we solve the equation of
state for three- and two-dimensional strongly interacting fermions
and find excellent agreement with experimental
{[}Science \textbf{335}, 563 (2012){]} and other theoretical results
above the temperature $0.5T_{F}$.
\end{abstract}


\maketitle
Using Feshbach resonances to tune the $s$-wave scattering length
of two-component atomic Fermi gases \cite{Bloch2008,Giorgini2008,Chin2010},
the experimental exploration of the crossover from Bose-Einstein condensates
(BEC) to Bardeen-Cooper-Schrieffer (BCS) superfluids in both three
(3D) \cite{Ketterle2008,Nascimbene2010,Horikoshi2010,Ku2012} and
two dimensions (2D) \cite{Martiyanov2010,Frohlich2011,Feld2011,Dyke2011,Makhalov2014,Murthy2015,Dyke2016,Fenech2016,Boettcher2016}
has attracted significant attention for understanding strongly interacting
phenomena. This has offered insight to other strongly interacting
systems such as high-$T_{c}$ superconductors \cite{Lee2006RMP},
nuclear matter \cite{Lee2006PRA} and quark-gluon plasma \cite{Kolb2004}.
These experiments have given a unique challenge to theorists as strongly
correlated matter cannot be quantitatively described by a simple mean-field
theory.

For strongly interacting systems, exact treatments only exist in some
limiting situations, for example, Bethe ansatz solutions in one dimension
\cite{Guan2013}, virial expansion at high temperature \cite{Liu2009,Leyronas2011,Rakshit2012,Liu2013}
and Tan relations at large momentum \cite{Tan2008a,Tan2008b,Tan2008c,Braaten2008}.
Numerically exact and sophisticated quantum Monte Carlo (QMC) simulations
have also been developed \cite{Astrakharchik2004,Bulgac2006,Houcke2012,Shi2015,Anderson2015},
however, these approaches have their own difficulty evaluating experimentally
relevant observables. In addition to the exact techniques, there have
been many attempts to solve strongly interacting fermions through
approximate diagrammatic theories. A commonly used approximation is
to sum over the complete geometric series of \emph{ladder} diagrams,
leading to the Gaussian pair fluctuation (GPF) theory \cite{Nozieres1985,SadeMelo1993,Hu2006,Liu2006,Parish2007,Diener2008,Watanabe2013,Marsiglio2015,He2015,Bighin2016}.
Though the GPF theory seems to provide consistent predictions for
recent experiments within certain errors \cite{Hu2008,Hu2010,Mulkerin2015},
it is hard to evaluate its validity. Improvements of using
fully dressed Green function in ladder diagrams, i.e., the partially
self-consistent pseudogap theory \cite{Chen2005} or fully self-consistent
$GG$ theory \cite{Haussman1994,Liu2005,Haussmann2007,Bauer2014},
meet similar problems. To develop a better strong-coupling theory
one needs to consider terms beyond GPF, which turns out to be a notoriously
difficult issue.

In this work, we attempt to tackle this daunting task and develop
a beyond GPF theory, as inspired by a dimensional $\epsilon$ expansion
\cite{NishidaPhDThesis,Nishida2006,Nishida2007a,Nishida2007b,Arnold2007}.
It was recognized that in the unitary regime \cite{Nussinov2006}
- where a bound state with zero energy appears - and near four dimensions
($d=4-\epsilon$), for small $\epsilon$, the two-component Fermi
gas behaves like a system of non-interacting composite bosons. This
is indicative of weaker effective interparticle interactions above
three dimensions, as characterized by the particle-particle vertex
function $\Gamma$. With such a re-interpretation of the small parameter
in the dimensional expansion, i.e., the use of $\Gamma$ instead of
$\epsilon$, we re-organize higher-order Feynman diagrams beyond GPF,
within the functional path-integral approach. In principle, the resulting
systematic expansion in terms of the vertex function is convergent
for dimensions where $\epsilon<1$, and may also asymptotically converge
at three dimensions, where $\epsilon=1$, following the extrapolation
strategy in Ref. \cite{Houcke2012}. Building upon this generalization
of the $\epsilon$ expansion, we examine the leading-order correction
to the GPF. As a test, we apply our theory to 3D and 2D strongly interacting
systems, finding excellent agreement with experimental benchmarks and other theoretical
techniques within a certain temperature window.

\textit{Effective field theory.} --- We consider the thermodynamic
potential, $\Omega=-k_{{\rm B}}T\ln\mathcal{Z}$, at a given temperature
$T$ using the functional path-integral formulation, which has been
extensively adopted in both 3D and 2D \cite{SadeMelo1993,Diener2008,He2015,Bighin2016}.
The partition function, $\mathcal{Z}=\int\mathcal{D}\left[\psi,\bar{\psi}\right]e^{-S\left[\psi,\bar{\psi}\right]}$,
where $\psi$ and $\bar{\psi}$ are independent Grassmann fields,
is defined through the action 
\begin{alignat}{1}
S[\psi,\bar{\psi}]=\int_{0}^{\hbar\beta}d\tau\int d\mathbf{r}\left[\sum_{\sigma}\bar{\psi}_{\sigma}(x)\partial_{\tau}\psi_{\sigma}(x)+H\right],
\end{alignat}
where the single-channel model Hamiltonian is $H=\sum_{\sigma}\bar{\psi}_{\sigma}(x)\mathcal{K}\psi_{\sigma}(x)+U_{0}\bar{\psi}_{\uparrow}(x)\bar{\psi}_{\downarrow}(x)\psi_{\downarrow}(x)\psi_{\uparrow}(x),$
$\mathcal{K}=-\hbar^{2}\nabla^{2}/(2m)-\mu$, $\beta=(k_{{\rm B}}T)^{-1}$,
$m$ is the mass of a fermion, $\mu$ is the chemical potential, and
throughout we shall use the notation $x=(\mathbf{x},\tau)$. We take
a contact attractive interaction, $U_{0}<0$, which has known divergences
and must be fixed \cite{Hu2010}. We will write most of the equations
detailed in this work using the bare interaction, dealing with the
divergences where necessary. Using the Hubbard-Stratonovich transformation
to write the action in terms of a bosonic field, $\Delta(q)$, and
expanding about its saddle point $\Delta_{0}$, $\Delta(q)=\Delta_{0}+\varphi_{q}$,
we take a perturbative expansion of the bosonic action in orders of
the fluctuation as $\scS_{{\rm eff}}\left[\Delta,\Delta^{*}\right]=\scS_{{\rm MF}}^{(0)}+\scS_{{\rm GF}}^{(2)}+\scS^{(3)}+\scS^{(4)}+\dots$,
where $\scS_{{\rm MF}}^{(0)}$ is the mean-field contribution and
the higher orders are 
\begin{alignat}{1}
\scS_{{\rm GF}}^{(2)} & =\frac{1}{2}\sum_{q}\left(\varphi_{q}^{*},\varphi_{-q}\right)\left[-\mathbf{\Gamma}^{-1}(q)\right]\twovec{\varphi_{q}}{\varphi_{-q}},\label{eq:all_the_S}\\
\scS^{(n)} & =\frac{1}{n}\textrm{Tr}\left[\left(\mathbf{G}_{0}(k)\mathbf{\mathbf{\Lambda}}(k)\right)^{n}\right].\label{eq:scS_largen}
\end{alignat}
Here we define $\mathbf{G}_{0}\mathbf{\mathbf{\Lambda}}=\mathbf{G}_{0}\varphi_{q}\sigma^{+}+\mathbf{G}_{0}\varphi_{-q}^{*}\sigma^{-}$,
where 
\begin{alignat}{1}
{\mathbf{G}_{0}}=\left(\begin{array}{cc}
i\omega_{m}-\xi_{\mathbf{k}} & -\Delta_{0}\\
-\Delta_{0} & i\omega_{m}+\xi_{\mathbf{k}}
\end{array}\right)^{-1},
\end{alignat}
$\xi_{\mathbf{k}}=\varepsilon_{\mathbf{k}}-\mu$ and $\varepsilon_{\mathbf{k}}=\mathbf{\hbar^{2}k}^{2}/(2m)$,
and $\sigma^{\pm}=\frac{1}{2}(\sigma_{1}\pm i\sigma_{2})$ are the
Pauli matrices. The trace in Eq.~(\ref{eq:scS_largen}) is over all
space and spin indices and we have used the summation convention $\sum_{k}\equiv(k_{B}T/V)\sum_{\mathbf{k},i\omega_{m}}$
and $\sum_{q}\equiv(k_{B}T/V)\sum_{\mathbf{q},i\nu_{n}}$, where $i\nu_{n}$
and $i\omega_{m}$ are bosonic and fermionic Matsubara frequencies,
respectively. The second-order Gaussian fluctuation term, $\scS_{{\rm GF}}^{(2)}$,
is the repeated scattering of two opposite spin fermions, and the
elements of the vertex function, $\Gamma(q)$, are given by, 
\begin{eqnarray}
\Gamma_{11}(q) & = & 1/U_{0}+\sum_{k}G_{11}^{(0)}(q-k)G_{11}^{(0)}(k),\\
\Gamma_{12}(q) & = & \sum_{k}G_{12}^{(0)}(q-k)G_{12}^{(0)}(k),
\end{eqnarray}
$\Gamma_{21}(q)=\Gamma_{12}(q)$ and $\Gamma_{22}(q)=\Gamma_{11}(-q)$.
In the normal phase where there is no superfluid order parameter,
i.e. $\Delta_{0}=0$, the vertex function is given by, $\Gamma^{-1}(q)=1/U_{0}+\sum_{k}G_{0}(q-k)G_{0}(k)$.
In the original Gaussian fluctuation theory (known alternatively as
the NSR theory) \cite{Nozieres1985}, terms beyond $\scS_{{\rm GF}}^{(2)}$
are simply discarded and the bosonic fields are integrated out, giving
$\Omega=\Omega_{0}+\Omega_{{\rm GF}}^{(2)}$, where $\Omega_{0}$
is the thermodynamic potential for a non-interacting Fermi gas and
the Gaussian fluctuation contribution is $\Omega_{{\rm GF}}^{(2)}=\sum_{q}\ln[-\Gamma^{-1}(q)].$
The number equation, $n=-\partial\Omega/\partial\mu$, should be satisfied
by adjusting the chemical potential for a given reduced temperature,
$T/T_{{\rm F}}$, and the equation of state can then be found. The
higher order terms, $\mathcal{S}^{(n)}$, contain beyond Gaussian
contributions of the bosonic fluctuation, $\varphi_{q}$, and the
treatment of these terms in the literature is sparse \cite{Gubbels2011}.

\begin{figure}
\includegraphics[width=0.48\textwidth]{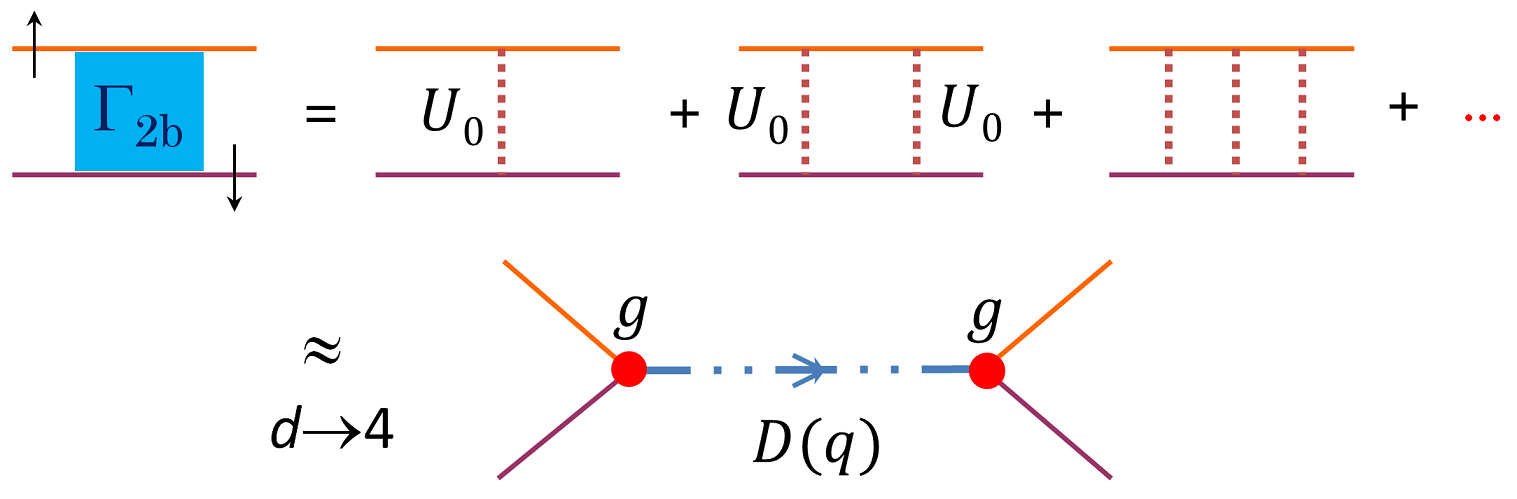} \caption{ Two fermion scattering in the unitary limit showing the repeated
ladder diagrams near four dimensions.}

\label{fig:graph} 
\end{figure}

\textit{Re-interpretation of the $\epsilon$ expansion}. --- To calculate
the beyond Gaussian contribution, we tie the GPF theory to the dimensional
$\epsilon$ expansion by clarifying the structure of the vertex function,
$\Gamma(q)$, near four dimensions \cite{SupplementalMaterial}. We
split the normal-state vertex function into its two- and many-body
parts, $\Gamma^{-1}(q)=\Gamma_{2b}^{-1}(q)+\Gamma_{mb}^{-1}(q)$,
where 
\begin{alignat}{1}
\Gamma_{2b}^{-1}(q)= & \frac{1}{U_{0}}-\sum_{\mathbf{k}}\frac{1}{i\nu_{n}-\varepsilon_{\mathbf{q}}/2-2\epsilon_{\mathbf{k}}-2\mu},\label{Eq:Gamma_2b}\\
\Gamma_{mb}^{-1}(q)= & \sum_{\mathbf{k}}\frac{f\left(\xi_{\mathbf{q}/2-\mathbf{k}}\right)+f\left(\xi_{\mathbf{q}/2+\mathbf{k}}\right)}{i\nu_{n}-\varepsilon_{\mathbf{q}}/2-2\varepsilon_{\mathbf{k}}-2\mu}.
\end{alignat}
For small $\epsilon=4-d$ and in the unitary limit, the two-body part
of the vertex function, $\Gamma_{2b}^{-1}(q)$, has a pole and dominates
the inverse vertex function, 
\begin{alignat}{1}
\Gamma_{2b}^{-1} & \simeq\frac{m^{2}}{8\pi^{2}\hbar^{4}\epsilon}\left(i\nu_{n}-\varepsilon_{\mathbf{q}}/2+2\mu\right)=\frac{1}{g^{2}D^{2}\left(q\right)},
\end{alignat}
where we define $g^{2}=(8\pi^{2}\hbar^{4}/m^{2})\epsilon$ and the
bosonic propagator, $D(q)=(i\nu_{m}-\varepsilon_{\mathbf{q}}+2\mu)^{-1}$.
Thus, we see that the vertex function,
$\Gamma(q)$, within the ladder approximation
 has the leading contribution of $\mathcal{O}(\epsilon)$
near four dimensions in the unitary regime. This is visualized in
Fig.~\ref{fig:graph}, where we show the contribution of the two-body
scattering near four dimensions. 

In the $\epsilon$ expansion, the series is arranged according to
orders of $\epsilon$, or equivalently $\Gamma_{2b}$ \cite{Nishida2006}.
While such an arrangement is convenient to analytically calculate
the next-to-leading order (NLO) \cite{Nishida2006} or next-to-next-to-leading
order (NNLO) of the expansion \cite{Arnold2007}, it was known that
one may encounter a convergence problem in dealing with some \emph{dangerous}
higher-order terms that contribute like $\mathcal{O}(n!\epsilon^{n})$
due to the exponentially large prefactor $n!$ \cite{NishidaAppendixA}.
These terms are contributions from the many-body part of the vertex
function, $\Gamma_{mb}$, to the Gaussian fluctuation part of the
thermodynamic potential, $\Omega_{{\rm GF}}^{(2)}$. The summation
of these terms is given by, $\sum_{n=1,q}^{\infty}(-1)^{n+1}\left[\Gamma_{2b}\Gamma_{mb}^{-1}\right]^{n}/n=\sum_{q}\ln\left[1+\Gamma_{2b}/\Gamma_{mb}\right]$
\cite{SupplementalMaterial}, and combining this with the two-body
part, $\sum_{q}\ln[-\Gamma_{2b}^{-1}(q)]$, we recover $\Omega_{{\rm GF}}^{(2)}$.
Therefore, since $\Omega_{{\rm GF}}^{(2)}$ contains one power of
the vertex function, which is $\mathcal{O}(\epsilon)$ at the NLO,
the $\epsilon$ expansion can be understood from the framework of the GPF theory.
This re-interpretation suggests that it might be more useful to make
an expansion in terms of the vertex function, $\Gamma(q)$, instead
of $\epsilon$.

\textit{Beyond GPF.} --- As previously noted we have the effective
bosonic action, $\scS_{{\rm eff}}$, and it is not possible to integrate
out the bosonic fluctuations for orders beyond $n=2$ without significant
approximations. Using the re-interpretation of the the $\epsilon$-expansion 
we expand the higher order action terms and
use the vertex function as a perturbation parameter, since the higher
order contributions to the action will contain multiples of the vertex
function and contribute $\mathcal{O}(\epsilon)$ near four dimensions.
In these cases we may treat the terms $\mathcal{S}^{(n>2)}$ as perturbative
terms with respect to $S_{{\rm GF}}^{(2)}$ and take them into account
order by order, using a standard diagrammatic approach. That is, by
denoting, $\hat{V}=\scS^{(3)}+\scS^{(4)}+...,$ we have for the partition
function 
\begin{alignat}{1}
\mathcal{Z} & =e^{-\mathcal{S}_{{\rm MF}}^{(0)}}\int\mathcal{D}\left[\Delta,\Delta^{*}\right]e^{-(\mathcal{S}_{{\rm GF}}^{(2)}+\hat{V})}\nonumber \\
 & =e^{-\mathcal{S}_{{\rm MF}}^{(0)}}\int\mathcal{D}\left[\Delta,\Delta^{*}\right]e^{-\mathcal{S}_{{\rm GF}}^{(2)}}\sum_{n=0}^{\infty}\frac{(-)^{n}}{n!}\left\langle \hat{V}_{1}\hat{V}_{2}\dots\hat{V}_{n}\right\rangle ,
\end{alignat}
where we have inserted a normalization term and defined, for any observable
(operator) $\hat{A}$, 
\begin{alignat}{1}
\left\langle \hat{A}\right\rangle =\frac{\int\mathcal{D}\left[\Delta,\Delta^{*}\right]\hat{A}\,e^{-\mathcal{S}_{{\rm GF}}^{(2)}}}{\int\mathcal{D}\left[\Delta,\Delta^{*}\right]e^{-\mathcal{S}_{{\rm GF}}^{(2)}}}.
\end{alignat}
Using the linked cluster expansion, we may write, 
\begin{alignat}{1}
\mathcal{Z}=e^{-\mathcal{S}_{{\rm MF}}^{(0)}}\int\mathcal{D}\left[\Delta,\Delta^{*}\right]e^{-\mathcal{S}_{{\rm GF}}^{(2)}}\exp\left(\sum_{l=1}^{\infty}U_{l}\right),
\end{alignat}
where the only contribution to the partition function is from the
differently-connected diagrams,
\begin{equation}
U_{l}=\frac{(-)^{l+1}}{l}\left\langle \hat{V}_{1}\hat{V}_{2}\dots\hat{V}_{l}\right\rangle _{{\rm dc}}.
\end{equation}
The expansion of the thermodynamic potential is then given by, 
\begin{equation}
\Omega=\Omega_{{\rm MF}}^{(0)}+\Omega_{{\rm GF}}^{(2)}+\frac{1}{\beta V}\sum_{l=1}^{\infty}\frac{(-)^{l+1}}{l}\left\langle \hat{V}_{1}\hat{V}_{2}\dots\hat{V}_{l}\right\rangle _{{\rm dc}}.
\end{equation}
We refer to Supplemental Material for discussions on how to calculate
diagrams related to $U_{l}$ \cite{SupplementalMaterial}. The expansion
should converge near four dimensions, where $\Gamma(q)$ is small.
Therefore, to select the important diagrams in the calculation of
the thermodynamic potential for 3D or 2D, 
we choose the leading diagrams in orders of $\epsilon$ near four dimensions.
Under this guidance, 
the leading order term beyond GPF is the connected diagram, $\langle{\rm S}^{(4)}\rangle$,
which for the normal state is given by, 
\begin{alignat}{1}
\frac{1}{\beta V}\langle{\rm S}^{(4)}\rangle=\sum_{k}\left[G_{0}(k)\Sigma_{0}(k)\right]^{2},
\end{alignat}
where the self-energy term is, $\Sigma_{0}(k)=\sum_{q}G_{0}(q-k)\Gamma(q)$ \cite{SupplementalMaterial}.
In the superfluid phase, it takes a more complicated form, $(\beta V)^{-1}\langle{\rm S}^{(4)}\rangle=\Omega^{(a)}+\Omega^{(b)}+\Omega^{(c)}$,
where 
\begin{alignat}{1}
\Omega^{(a)}= & \sum_{k}  G_{11}^{(0)}  (k)G_{11}^{(0)}(k)\Sigma_{11}^{(0)}(k)\Sigma_{11}^{(0)}(k),\label{eq:S4a}\\
\Omega^{(b)}= & \sum_{k}  G_{12}^{(0)}  (k)G_{12}^{(0)}(k)\Sigma_{11}^{(0)}(k)\Sigma_{22}^{(0)}(k),\label{eq:S4b}\\
\Omega^{(c)}= & \sum_{kq_{1}q_{2}}  G_{12}^{(0)}  (k)G_{11}^{(0)}(k-q_{1})G_{12}^{(0)}(k-q_{1}+q_{2})\nonumber \\
 & \hspace{21pt}G_{22}^{(0)}  (k+q_{2})\Gamma_{11}(q_{1})\Gamma_{11}(q_{2}),\label{eq:S4c}
\end{alignat}
$\Sigma_{11}^{(0)}(k)=\sum_{k}G_{11}^{(0)}(q-k)\Gamma_{11}^{(0)}(k)$
and $\Sigma_{22}^{(0)}(k)=-\Sigma_{11}^{(0)}(-k)$. As we can see,
$\langle{\rm S}^{(4)}\rangle$ contains two $\Gamma(q)$ and near
four dimensions is the NNLO, $\mathcal{O}(\epsilon^{2})$ contribution.
In other words, if we calculate $\langle{\rm S}^{(4)}\rangle$ in
$d=4-\epsilon$ dimensions and numerically extract the coefficients
at $\mathcal{O}(\epsilon^{2})$, we are able to recover the NNLO $\epsilon$
expansion \cite{Arnold2007}. Higher order $\epsilon$ expansions can
be obtained if we go further beyond GPF.

\begin{figure}
\includegraphics[width=0.48\textwidth]{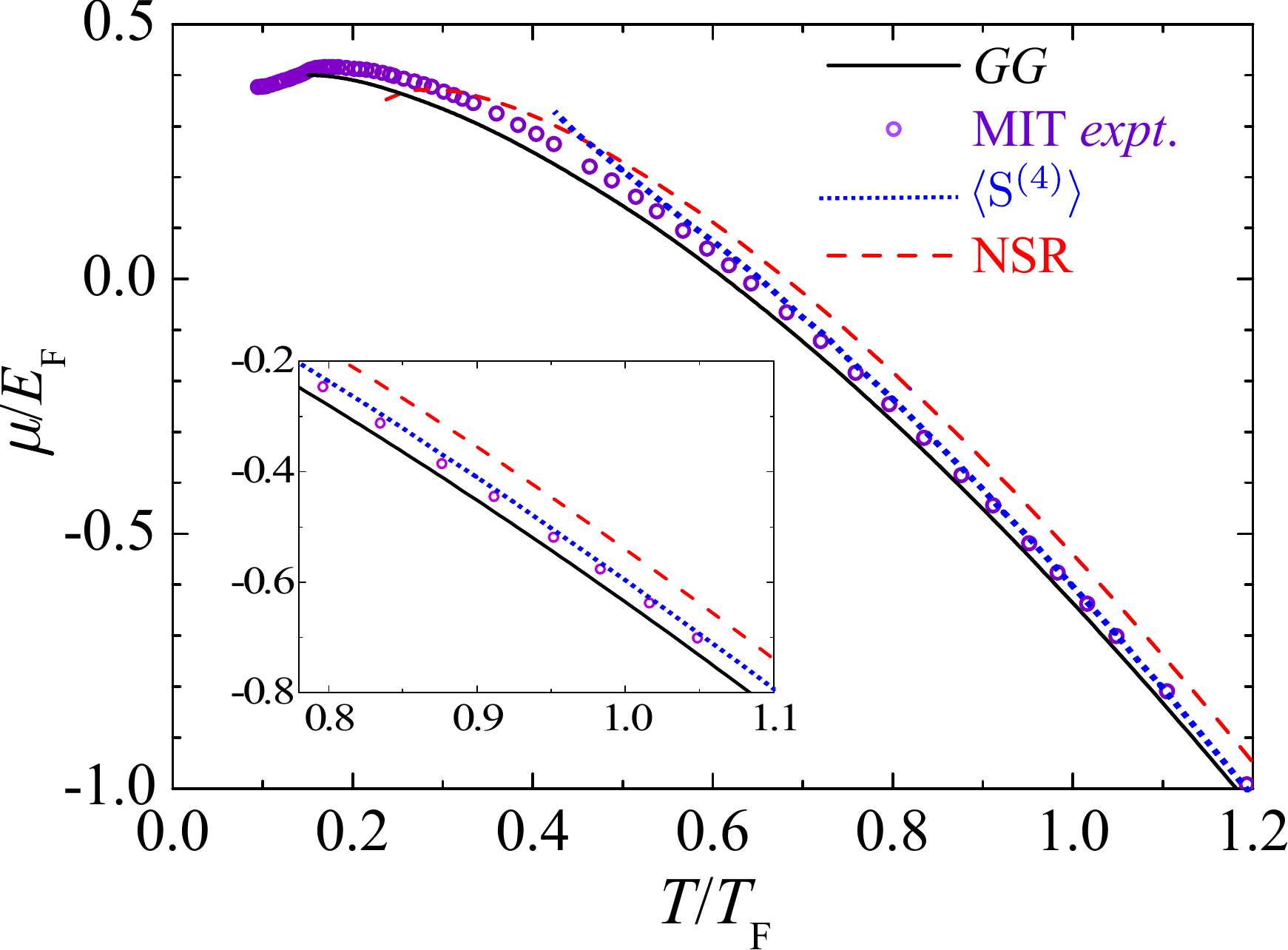}\caption{The chemical potential, $\mu/E_{{\rm F}}$, plotted as a function
of reduced temperature, $T/T_{{\rm F}}$, found from the MIT experiment
(maroon circles) \cite{Ku2012}, $GG$ $T$-matrix (black solid) \cite{Haussman1994},
$\textrm{NSR}$ (red dashed) \cite{Nozieres1985} and $\langle{\rm S}^{(4)}\rangle$
(blue dotted). The inset highlights the high-temperature region for
the same figure.}

\label{fig:chem_3D} 
\end{figure}

\textit{Equation of state.} --- As an application of our beyond GPF
theory, we determine the equation of state of an above threshold strongly interacting
Fermi gas in both 3D and 2D, by solving the number equation for the
thermodynamic potential, $\Omega=\Omega_{{\rm MF}}^{(0)}+\Omega_{{\rm GF}}^{(2)}+(\beta V)^{-1}\langle{\rm S}^{(4)}\rangle$.
Material \cite{SupplementalMaterial}. 
We use $\langle{\rm S}^{(4)}\rangle$
as the shorthand notation for our beyond GPF theory in the normal
state.

In Fig.~\ref{fig:chem_3D}, we report the chemical potential of a
3D unitary Fermi gas as a function of reduced temperature, $T/T_{{\rm F}}$,
predicted from the $\langle{\rm S}^{(4)}\rangle$, ${\rm NSR}$ \cite{Nozieres1985}
and self-consistent $GG$ theories \cite{Haussman1994}, and measured
by the MIT experiment \cite{Ku2012}. Here, we find that the $\langle{\rm S}^{(4)}\rangle$
prediction is in excellent agreement with the experimental data down
to $0.5T_{F}$, below which our prediction begins to diverge away
from the ${\rm NSR}$ solution and pairing fluctuations in $\langle{\rm S}^{(4)}\rangle$
start to dominate. This is because, as the temperature becomes lower
the effective interaction and fluctuation between pairs become stronger
and the NSR approximation itself \cite{Liu2006,Parish2007,Hu2008},
and leading-order correction beyond GPF are no longer controllable.

\begin{figure}
\includegraphics[width=0.48\textwidth]{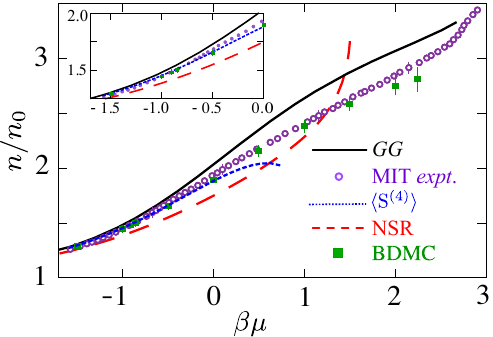}\caption{The density equation of state, $n/n_{0}$, plotted as a function of
$\beta\mu$, normalized by the ideal gas result, $n_{0}$, at the same
temperature $T$ and chemical potential $\mu$, found from the MIT
experiment (maroon circles with error bars), $GG$ (black solid),
${\rm NSR}$ (red dashed), $\langle{\rm S}^{(4)}\rangle$ (blue dotted),
and bold-diagrammatic QMC (green squares) \cite{Houcke2012}. The
inset blows up the non-degenerate region for the same plot. }

\label{fig:dens3D} 
\end{figure}

A further comparison is given in Fig.~\ref{fig:dens3D}, where we
plot the density equation of state, $n/n_{0}$, as a function of $\beta\mu$,
normalized by the ideal gas result at the same temperature $T$ and
chemical potential $\mu$. We find again an excellent agreement between
the $\langle{\rm S}^{(4)}\rangle$ prediction and experimental data
up to values of $\beta\mu\simeq0$, as seen clearly in the inset of
Fig.~\ref{fig:dens3D} for high temperatures. The $\langle{\rm S}^{(4)}\rangle$
prediction agrees well with the experimental data, and
shows that our beyond GPF theory is and improvement on $GG$ theory and
comparable to bold-diagrammatic QMC \cite{Houcke2012}, up to
slightly below the Fermi degeneracy. The $\langle{\rm S}^{(4)}\rangle$
theory breaks down at lower temperatures, $\beta\mu\simeq1$, and
as mentioned earlier this is due to the unphysical pair fluctuations
dominating as we only calculate the leading term. Our calculations
are not stable towards the experimentally measured critical temperature.
To understand the superfluid transition using our beyond GPF theory,
a below $T_{c}$ calculation with the inclusion of a superfluid order
parameter will be implemented and reported later in a more detailed
publication.

\begin{figure}
\includegraphics[width=0.48\textwidth]{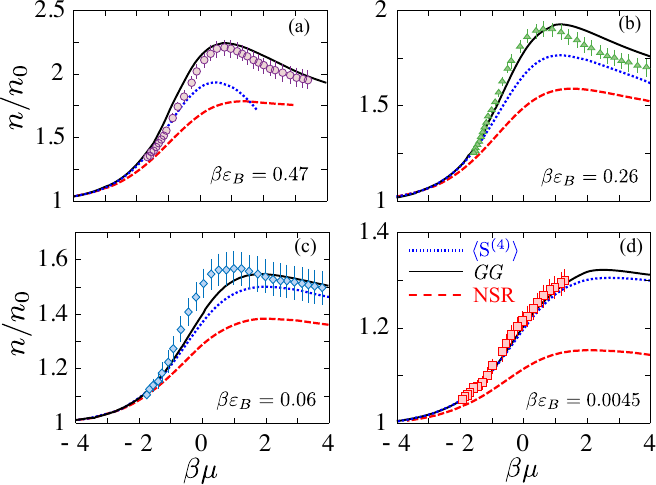} \caption{The density equation of state, $n/n_{0}$, normalized by the ideal
gas result at the same temperature for the $GG$ (black solid) \cite{Bauer2014,Mulkerin2015},
${\rm NSR}$ (red dashed) \cite{Marsiglio2015,Mulkerin2015}, $\langle{\rm S}^{(4)}\rangle$
and the experimental results \cite{Fenech2016} at interaction strengths
$\beta\varepsilon_{B}=0.47$, $\beta\varepsilon_{B}=0.26$ , $\beta\varepsilon_{B}=0.06$
and $\beta\varepsilon_{B}=0.0045$. }

\label{fig:dens2D} 
\end{figure}

Encouraged by the excellent agreement between the beyond GPF theory
and experiment in 3D, we now turn to consider the density equation
of state for a 2D system, where pair fluctuations are believed to
become larger. The results are shown in Fig.~\ref{fig:dens2D}
for the $\langle{\rm S}^{(4)}\rangle$ (blue dotted) and for comparison
we plot two theoretical approaches, the NSR (red dashed) \cite{Marsiglio2015,Mulkerin2015}
and $GG$ (black solid) calculation \cite{Bauer2014,Mulkerin2015},
and the experimental results \cite{Fenech2016} for interaction strengths
$\beta\varepsilon_{B}=0.47$ (purple circles), $\beta\varepsilon_{B}=0.26$
(green triangles), $\beta\varepsilon_{B}=0.06$ (blue diamonds), and
$\beta\varepsilon_{B}=0.0045$ (red squares).

The addition of the higher order terms in the $\langle{\rm S}^{(4)}\rangle$
calculation greatly improves the NSR theory for all interactions and
temperatures, and the resulting prediction is comparable to the experimental
data and $GG$ calculation up to $\beta\mu\simeq0$. We can see the
$\langle{\rm S}^{(4)}\rangle$ breaking down for low temperatures
in Fig.~\ref{fig:dens2D}(a) and this is due to the fluctuation of
pairs becoming more important as in 3D. For the weaker interactions
in Figs.~\ref{fig:dens2D}(c) and \ref{fig:dens2D}(d), the inclusion
of the $\langle{\rm S}^{(4)}\rangle$ calculation approaches
the experimental and $GG$ results.

\textit{Conclusions.} --- We have extended the many-body strong-coupling
theory beyond the commonly used Gaussian fluctuation approximation
(i.e., the NSR \cite{Nozieres1985} or GPF theory \cite{Hu2006}).
Inspired by the dimensional expansion near four dimensions \cite{Nishida2006}
and using the functional path-integral formulation of the thermodynamic
potential \cite{SadeMelo1993}, we artificially treat a strongly interacting
Fermi gas as a system of weakly interacting Cooper pairs and use the
vertex function summed over the ladder-type diagrams as a small parameter.
This treatment is well justified near four dimensions, where the vertex
function is indeed small \cite{Nishida2006}. Following this generalization
of the dimensional expansion, we re-organize the Feynman diagrams
and determine the leading correction term to the Gaussian fluctuations.
Applying such a beyond Gaussian fluctuation theory to the three-dimensional
strongly interacting unitary Fermi gas in its normal state, we have
calculated the equation of state and compared the prediction with the
latest experimental data \cite{Ku2012} and other theoretical results
\cite{Nozieres1985,Haussman1994}. We have found a sizable improvement
on previous many-body calculations down to temperatures of $0.5T_{{\rm F}}$.
To further examine the advantage of the theory, we have considered
a strongly interacting two-dimensional Fermi gas, for which the pair
fluctuations are more significant, and have shown that our theory
significantly improves the NSR calculation and captures the high-temperature
behavior for strong interactions. 

Our theory for the normal state breaks down before the superfluid
transition in both 3D and 2D as the NSR theory itself breaks down
\cite{Liu2006,Parish2007}, and we expect that the addition of more
terms will lower the temperature range of validity. At zero temperature,
where the GPF theory is more reliable \cite{Hu2006,Diener2008,He2015},
we anticipate our theory (i.e., Eqs.~(\ref{eq:S4a})-(\ref{eq:S4c}))
will provide quantitatively accurate predictions for strongly interacting
Fermi gases.
\begin{acknowledgments}
We would like to thank Lianyi He and Xu-Guang Huang for useful discussions
and Felix Werner for sending us the bold diagrammatic QMC data. XJL
and HH acknowledge the support from the ARC Discovery Projects (FT130100815,
DP140100637, DP140103231 and FT140100003).\end{acknowledgments}

\end{document}